\pdfoutput=1

\documentclass[12pt]{article}

\usepackage{hyperref}

\usepackage[usenames]{color}
\usepackage{lscape}
\usepackage{amssymb}
\usepackage{rotating}
\usepackage[T1]{fontenc}
\usepackage{graphicx}
\usepackage{fancyhdr}
\usepackage{amsmath}
\usepackage{ae}
\usepackage{aecompl}
\usepackage{cite}

\usepackage[tight]{subfigure}

\usepackage{xcomment}

\newcommand{\be}{\begin{equation}}
\newcommand{\ee}{\end{equation}}
\newcommand{\ba}{\begin{eqnarray}}
\newcommand{\ea}{\end{eqnarray}}

\newcommand{\lsim}
{\;\raisebox{-.3em}{$\stackrel{\displaystyle <}{\sim}$}\;}
\newcommand{\gsim}
{\;\raisebox{-.3em}{$\stackrel{\displaystyle >}{\sim}$}\;}

\newlength{\dinwidth}
\newlength{\dinmargin}
\begin{document}

\thispagestyle{empty}

\begin{flushright}
IPPP/16/17
\end{flushright}

\vspace*{15mm}

\centerline{\Large\bf Interpreting 750 GeV Diphoton Excess in Plain NMSSM} 

\vspace*{15mm}

\centerline{\bf M.~Badziak${}^{a,b}$,  M.~Olechowski${}^a$,  S.~Pokorski${}^a$ and K.~Sakurai${}^c$}
\vspace*{5mm}

\centerline{${}^a$\em Institute of Theoretical Physics,
Faculty of Physics, University of Warsaw} 
\centerline{\em ul.~Pasteura 5, PL--02--093 Warsaw, Poland} 
\centerline{${}^b$\em Berkeley Center for Theoretical Physics, Department of Physics,}
\centerline{\em and Theoretical Physics Group, Lawrence Berkeley National Laboratory,}
\centerline{\em University of California, Berkeley, CA 94720, USA}
\centerline{${}^c$\em Institute for Particle Physics Phenomenology, 
Department of Physics, } 
\centerline{\em University of Durham, Science Laboratories, South Road, Durham, DH1 3LE, UK}

\vskip 1cm

\centerline{\bf Abstract}
\vskip 3mm
NMSSM has enough ingredients to explain the diphoton excess at 750 GeV: singlet-like (pseudo) scalar 
($a$) $s$ and higgsinos as heavy vector-like fermions.
We consider the production of the 750 GeV singlet-like pseudo scalar $a$ from a decay of the doublet-like pseudo scalar $A$, and  the  subsequent
decay  of $a$ into two photons via higgsino loop.
We demonstrate that 
this cascade decay of the NMSSM Higgs bosons can explain the diphoton excess at 750 GeV.

\vskip 3mm

\newpage

\section{Introduction}

Recently ATLAS and CMS have reported excesses in the diphoton mass distribution around $m_{\gamma \gamma} \simeq 750$ GeV
in their 13 TeV data.
The local significance assuming narrow width is $\sim 3.6 \, \sigma$ for ATLAS \cite{ATLAS-CONF-2015-081} 
and $\sim 2.6 \, \sigma$ for CMS \cite{CMS:2015dxe}.
ATLAS and CMS have presented their updated analyses at Moriond conference.
With the improved analyses, the local significance has increased to $\sim 3.9 \, \sigma$
and $\sim 3.4 \, \sigma$ for ATLAS and CMS, respectively \cite{ATLAS-CONF-2016-018, CMS:2016owr}.
Fitting that excess with a narrow resonance around 750 GeV, CMS  reports for 
the cross section times branching ratio, $\sigma_{13 \rm TeV} \cdot {\rm BR}_{\gamma \gamma}$,
the value $2.6 \div 7.7$ fb   at 1\,$\sigma$  and  $0.85\div (11-12)$fb
at 2\,$\sigma$ (see Fig.10 of \cite{CMS:2016owr}). The CMS fit of the excess around $m_{\gamma \gamma} \simeq 750$ GeV in the 8 TeV data gives $0.31 \div 1.00$ fb at 1-$\sigma$ and $0.06\div
1.45$ fb at 2-$\sigma$ \cite{CMS:2016owr}\footnote{
    In Fig.10 of \cite{CMS:2016owr}, CMS rescaled the fitted cross section of the 8 TeV result to 13 TeV assuming the $gg$ initial state.
    We rescale this back into 8 TeV.   
}. The ATLAS collaboration has not provided such a detailed analysis for a narrow resonance hypothesis. A fit   reported in ref.~\cite{Franceschini:2016gxv}  gives for $\sigma \cdot {\rm BR}_{\gamma \gamma}$ the values
$\simeq 4 \div 7$ fb   and  $\simeq 0 \div 0.42$ fb at 1\,$\sigma$ at 13 TeV and 8 TeV, respectively. No information about the 2\,$\sigma$ regions is
available.

The possible interpretation and implications of the excess has been intensively studied.
Most such studies introduce new particles to account for the excess without asking about their UV origin,
and interpretation within the known models in particular 
Minimal Supersymmetric Starndard Model (MSSM) and Next-to-Minimal Supersymmetric Standard Model (NMSSM)
is rare.\footnote{For
    R-parity violating (RPV) MSSM see \cite{Allanach:2015ixl, Ding:2015rxx} and for  
    NMSSM with $pp \to H \to aa \to (\gamma \gamma)(\gamma \gamma)$ see \cite{Ellwanger:2016qax, Domingo:2016unq}.
}
In this paper we study the possibility to explain the diphoton excess within the framework of NMSSM
without introducing additional particles.

One of the most straightforward interpretations of the excess is to consider a direct production of a scalar or pseudoscalar 750 GeV particle, $X$, decaying to two photons: 
$\alpha \beta \to X \to \gamma \gamma$, where $\alpha$, $\beta$ are the initial state partons.
If the model is renormalizable, $X \to \gamma \gamma$ suggests the existence of
electrically-charged 
vector-like fermions (or scalars) coupled to $X$ 
\cite{Franceschini:2015kwy,McDermott:2015sck,
Ellis:2015oso,Gupta:2015zzs,
Martinez:2015kmn,Fichet:2015vvy,
Bian:2015kjt,Falkowski:2015swt,
Bai:2015nbs,Dhuria:2015ufo,
Chakraborty:2015jvs,Wang:2015kuj,
Hernandez:2015ywg,Huang:2015rkj,
Badziak:2015zez,Cvetic:2015vit,
Cheung:2015cug,Zhang:2015uuo,
Hall:2015xds,Wang:2015omi,
Salvio:2015jgu,Son:2015vfl,
Cai:2015hzc,Bizot:2015qqo,
Hamada:2015skp,Kang:2015roj,
Jiang:2015oms,Jung:2015etr,
Gu:2015lxj,Goertz:2015nkp,
Ko:2016lai,Palti:2016kew,
Karozas:2016hcp,Bhattacharya:2016lyg,
Cao:2016udb,Faraggi:2016xnm,
Han:2016bvl,Kawamura:2016idj,
King:2016wep,Nomura:2016rjf,
Harigaya:2016pnu,Han:2016fli,
Hamada:2016vwk,Bae:2016xni,
Salvio:2016hnf,
Barbieri:2016cnt,
Pilaftsis:2015ycr,
Dev:2015isx,
Dev:2015vjd,
Arbelaez:2016mhg,
Hernandez:2016rbi,
Hernandez:2015hrt,
Ge:2016xcq,
Chao:2016aer,
Chao:2016mtn,
Han:2015qqj,
Han:2016bus,
Anchordoqui:2015jxc,
Harigaya:2015ezk
}, which generate the effective operator $X F^{\mu \nu} F_{\mu \nu} (\tilde F_{\mu \nu})$. 
Such fermions should be heavier than $m_X / 2 \simeq 375$ GeV, otherwise the diphoton rate is strongly suppressed
because $X$ predominantly decays into the vector-like fermions on shell.
Similar argument disfavours the possibility to identify $X$ as the heavy higgs bosons in the MSSM or 2HDM,\footnote{See however
\cite{Djouadi:2016ack,Djouadi:2016oey}.}
because in such models $X$ predominantly decays into $t \bar t$ and/or $b \bar b$ \cite{Angelescu:2015uiz}.
In general, in such scenarios the decay branching ratios of $X$ are strongly correlated with the production cross section.

Another possibility is to consider the production of $X$ from a decay of a heavy resonance $Y_r$
associated with another particle $Y_d$: $\alpha \beta \to Y_r \to Y_d X$, $X \to \gamma \gamma$ 
\cite{Franceschini:2015kwy,Huang:2015evq,Altmannshofer:2015xfo,Bi:2015lcf,Ding:2016udc}.
This topology has two advantages.
First, ${\rm BR}(X \to \gamma \gamma)$ is independent of the production cross section of the resonance.
This is not the case for the previous topology, 
because a large production cross section leads to a large rate of the inverse decay process $X \to \alpha \beta$,
which suppresses ${\rm BR}(X \to \gamma \gamma)$.
Second, the mass of $Y_r$ has to be larger than $m_X \simeq 750$ GeV,
and the 13 TeV production cross section of $Y_r$ is more enhanced with respect to the 8 TeV cross section,
compared to the previous topology. In this context  we notice that, while there is no big tension between 8 and 13 TeV data  in the CMS fits
interpreted as a direct production of a  750 GeV resonance, 
the fit of ref.~\cite{Franceschini:2016gxv} to the ATLAS data shows
such a tension   well above 2\,$\sigma$ level. For instance,
if the initial partons are gluons, $\alpha\beta = gg$,  translating the results of that fit for 8 TeV, interpreted as a direct production of the 750
GeV resonance,  to 13 TeV  clearly shows the problem. 
Thus, the cascade topology may slightly  help to  reconcile the ATLAS data  at 8 and 13 TeV and  the results of both experiments.

\begin{figure}
    \centering 
    \includegraphics[width=0.45\textwidth]{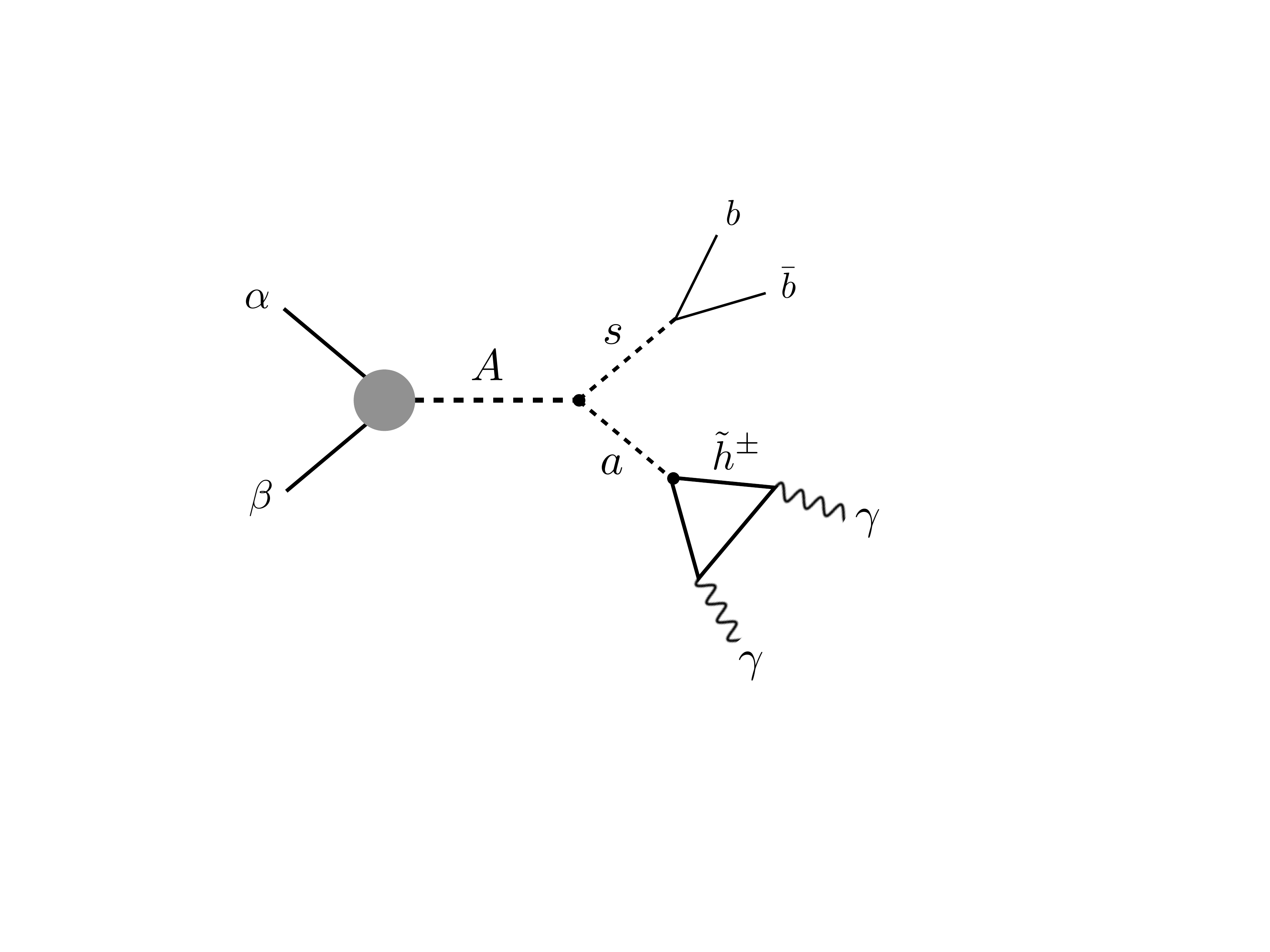}
    \vspace{-0.4cm}
\caption{\small 
An NMSSM Higgs boson cascade decay contributing to the diphoton excess.
The $\alpha$ and $\beta$ denote the initial state partons.
If $(\alpha, \beta) = (b \bar b)$, one also expects extra $b$ jets in the forward region.
\label{fig:diagram}}
\end{figure}

This topology can be relatively easily realised in the NMSSM by identifying 
$Y_r = A$, $Y_d = s$ and $X = a$: 
$\alpha \beta \to A \to s a$, $a \to \gamma \gamma$, as shown in Fig.~\ref{fig:diagram}, 
where $A$ is the doublet-like pseudo scalar and ($a$) $s$ is the singlet-like (pseudo) scalar.
In NMSSM $a \to \gamma \gamma$ is induced by a higgsino loop diagram also shown in Fig.~\ref{fig:diagram}.
The $Y_d = h$ is  disfavoured because  non-zero $Aha$ coupling requires doublet-singlet mixing in the pseudo-scalar sector ($Aa$ mixing), suppressing
$a \to \gamma \gamma$ branching ratio.
In our scenario, $s$ predominantly decays into $b \bar b$ through a mixing with $H$.
Although the current data would not have enough sensitivity to discriminate 
these extra jets from other jets with QCD origin,
this scenario can be tested by looking at these $b$-jets in the future analysis. 

The paper is organised as follows.
In section~\ref{sec:pure} we demonstrate our scenario in a simplified framework in which 
the mixing between singlet and doublet states are ignored.
In section~\ref{sec:nmssm} we consider
how our scenario can be realised in the NMSSM taking the effect of mixing into account.
We conclude this paper in section~\ref{sec:concl}.

\section{Interpretation with pure states}
\label{sec:pure}

We first 
discuss our scenario
in a simplified framework where
the resonance $A$ is pure doublet state and the lightest CP even and odd Higgs bosons, $s$ and $a$, 
are exclusively originated from the singlet field $S$.
The signal of the diphoton excess  is given by
\ba
(\sigma \cdot {\rm BR})^{\rm signal}\equiv
\sigma(pp \to A) \cdot {\rm BR}(A \to s a) \cdot {\rm BR}(a \to \gamma \gamma) 
\label{eq:13tevfit}
\ea
where the cross section $\sigma(pp \to A) $ depends on the centre of mass energy of the proton-proton collision.

\begin{figure}
    \centering 
    \includegraphics[width=0.49\textwidth]{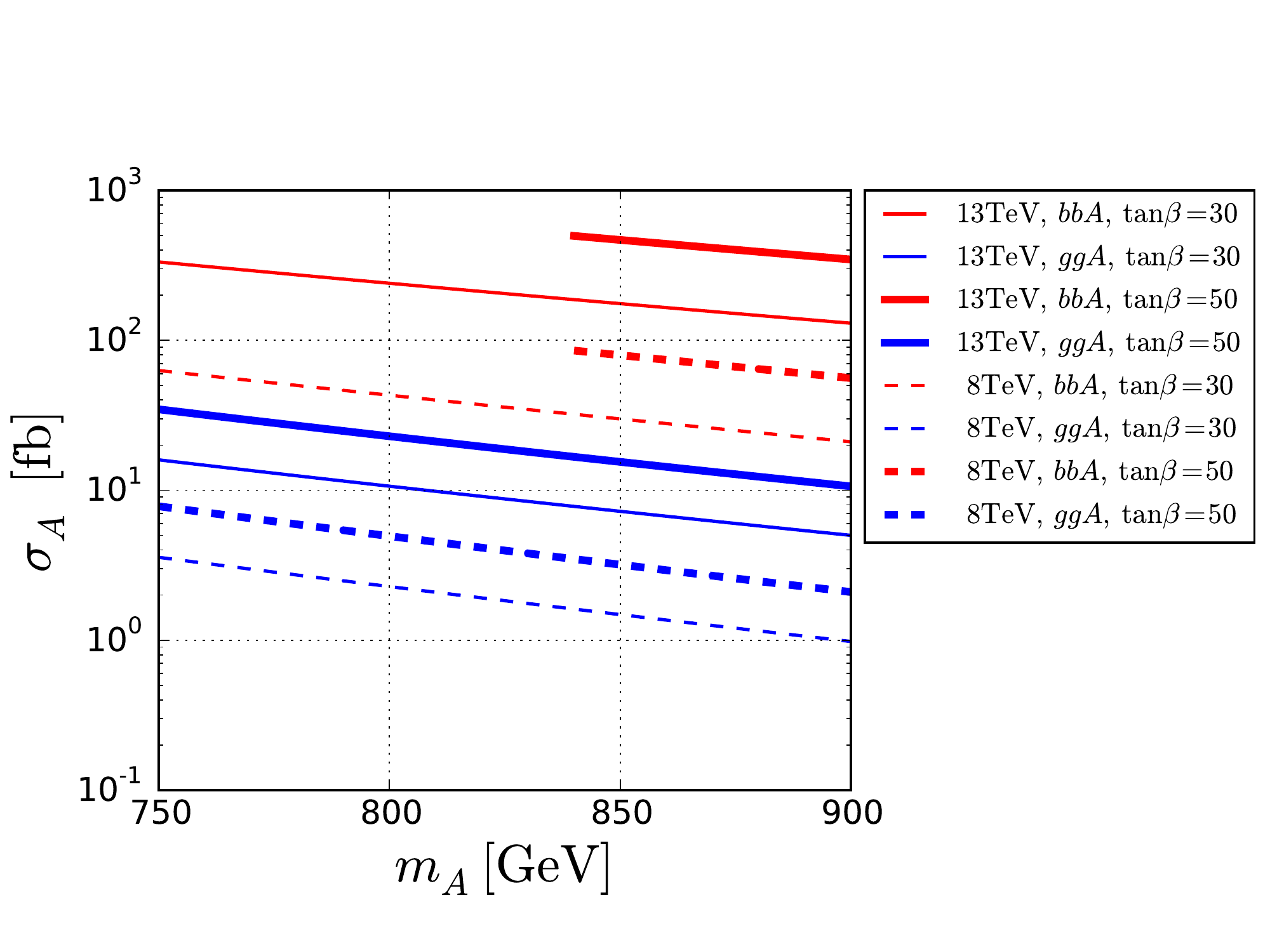}    
    \includegraphics[width=0.49\textwidth]{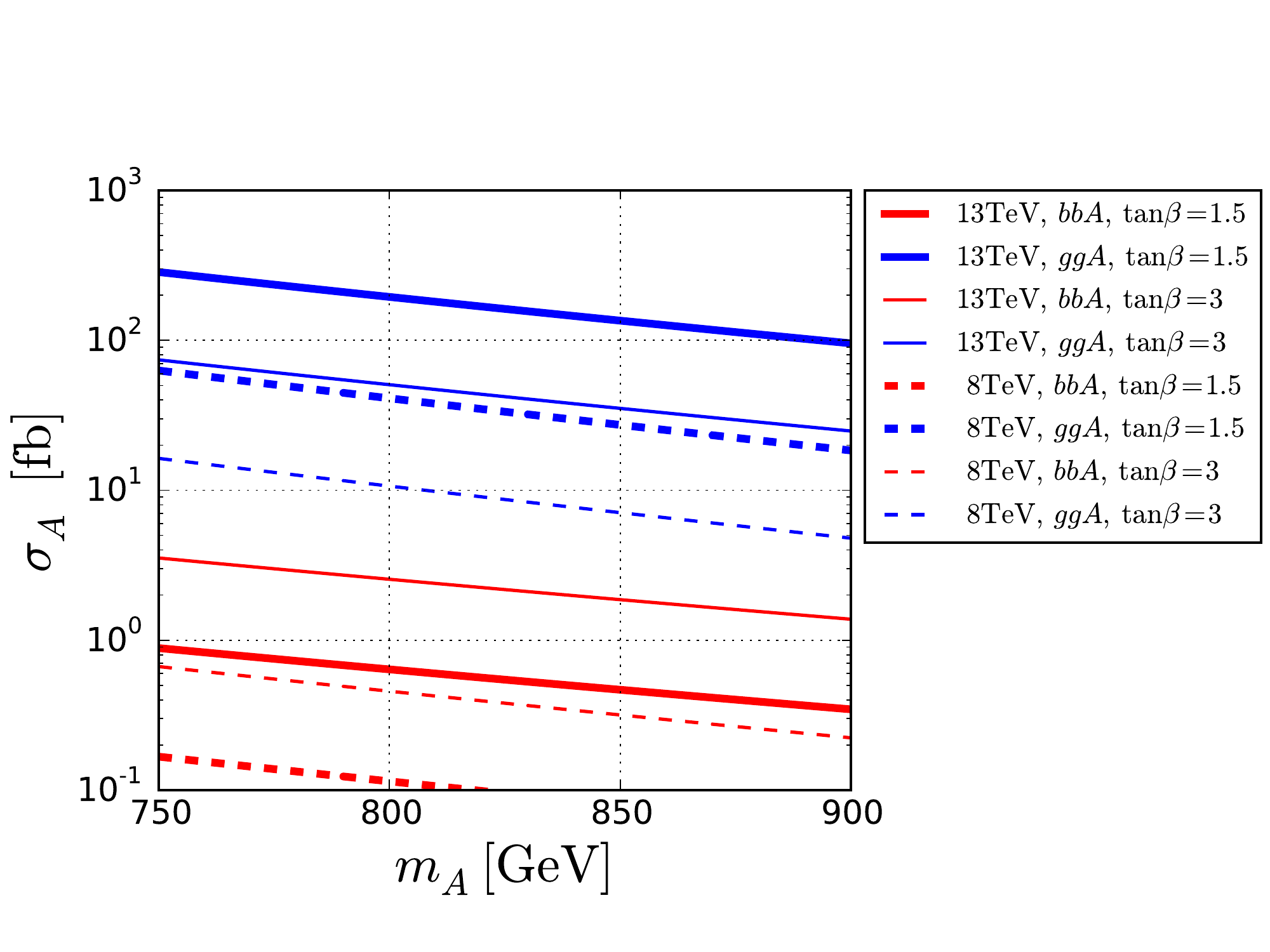}       
    \vspace{-0.4cm}
\caption{\small
Production cross section of $A$ from the $b \bar b$ (red) and $gg$ (blue) initial states 
as a function of $m_A$ for $\sqrt{S} = 13$ (solid) and 8 (dashed) TeV.
In the left (right) panel, the thick and thin lines correspond to $\tan \beta = 50$ and 30 (1.5 and 3),
respectively.
For the $b \bar b$ initial state with $\tan\beta = 50$, 
$m_A \lesssim 840$ GeV is excluded by the $b \bar b \to A \to \tau^+ \tau^-$ search.
\label{fig:xsec}}
\end{figure}

Fig.~\ref{fig:xsec} shows the NLO production cross section of $A$ from the  $b \bar b$ (red) and $gg$ (blue) initial states 
as a function of $m_A$ for $\sqrt{S} = 13$ (solid) and 8 (dashed) TeV.
In the left (right) panel of Fig.~\ref{fig:xsec}, 
the thick and thin lines correspond to $\tan \beta = 50$ and 30 (1.5 and 3), respectively.
The cross sections are calculated using {\tt SusHi v.1.5.0} \cite{Harlander:2012pb, Harlander:2003ai, Aglietti:2004nj,Bonciani:2010ms,
Degrassi:2010eu,Degrassi:2011vq,
Degrassi:2012vt,Harlander:2005rq}.
As can be seen, the $\sqrt{S} = 13$ TeV production cross section for large (small) $\tan\beta$ values is dominated by $b \bar b$  ($gg$) initial state. 
It  can be as large as 400 (200) fb for $\tan \beta = 50$ (30) at $m_A \sim 850$ GeV. 
The cross section enhances from 8 TeV to 13 TeV by factor of 
5 for $gg$ and 6.7 for $b \bar b$ initial states.
We impose the $m_A$ dependent upper limit on $\sigma(b \bar b \to A) \cdot {\rm BR}(A \to \tau^+ \tau^-)$
and $\sigma(gg \to A) \cdot {\rm BR}(A \to \tau^+ \tau^-)$
obtained from the 8 TeV CMS search for the neutral Higgs boson decaying to di-tau \cite{CMS:2015mca}.
We found the $m_A \lsim 840$ GeV is excluded by this constraint 
for $b \bar b$ initial state at $\tan \beta = 50$
and this region is not shown in Fig.~\ref{fig:xsec}.
For $\tan\beta = 30$ or the $gg$ initial state, the whole region 
with $m_A > 750$ GeV is allowed.

We define the interaction between $A$-$s$-$a$ as
\be
{\cal L} \supset g_{Asa} A s a \,.
\ee
With the coupling $g_{Asa}$, the partial decay rate of $A \to sa$ is given by 
\be
\Gamma(A \to s a) = \frac{|g_{Asa}|^2}{16 \pi m_A} \bar \lambda( \frac{m_s^2}{m_A^2}, \frac{m_a^2}{m_A^2}),
\label{eq:G_Asa}
\ee
where $\bar \lambda(a, b) \equiv 1 + a^2 + b^2 - 2 (a + b + ab)$.
In what follows we assume $m_s = 65$ GeV and 
$815 \le m_A \le 875$ GeV.
In this parameter region, $h \to ss$ and $A \to h a$ are kinematically 
forbidden 
while $A \to s a$ is allowed.

\begin{figure}
    \centering 
    \includegraphics[width=0.6\textwidth]{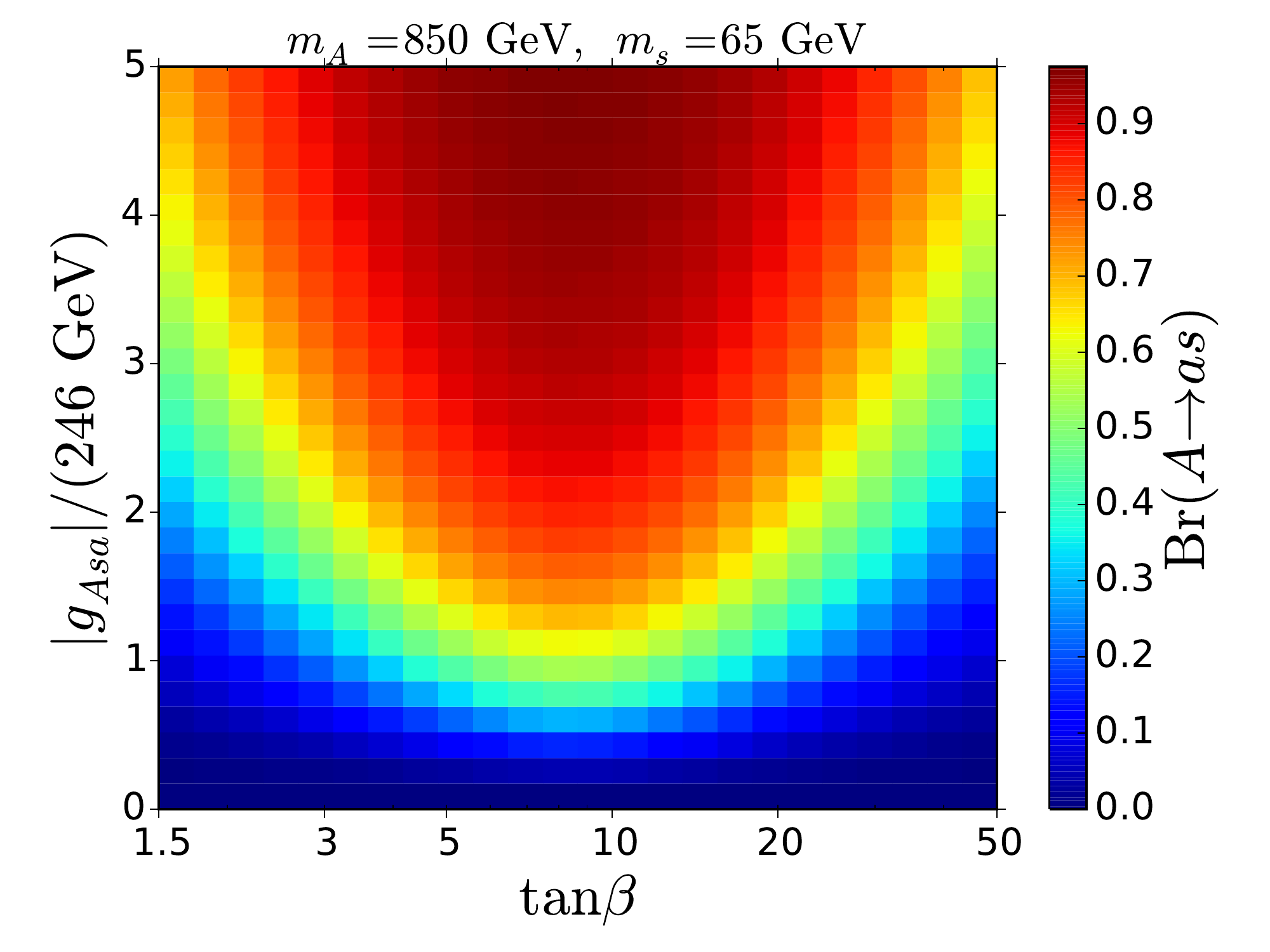}       
    \vspace{-0.4cm}
\caption{\small
Branching ratio of $A \to sa$
as a function of $\tan\beta$ and $g_{Asa}/(246 \, \rm GeV)$. 
We fix $m_A = 850$ GeV and $m_s = 65$ GeV.
\label{fig:br}}
\end{figure}

The $A \to sa$ decay mode competes with $A \to b \bar b$ and $A \to t \bar t$ in the large and small $\tan\beta$ regimes, respectively.  
The partial decay rates are given by
\be
\Gamma(A \to b \bar b) = \frac{3 \alpha_W m_A}{8 m_W^2} m_b^2 \tan^2\beta \Big( 1 - \frac{4 m_b^2}{m_A^2} \Big)^{1/2}, ~~~
\Gamma(A \to t \bar t) = \frac{3 \alpha_W m_A}{8 m_W^2} m_t^2 \cot^2\beta \Big( 1 - \frac{4 m_t^2}{m_A^2} \Big)^{1/2} \,.
\label{eq:Aff}
\ee
The decay modes into gauge bosons are highly suppressed due to the 
CP property.
Fig.~\ref{fig:br} shows the branching ratio of $A \to sa$ for $m_A = 850$ GeV, $m_s = 65$ GeV as a function of $\tan\beta$ and $|g_{Asa}|/(246 \, \rm GeV)$.
At a fixed $g_{Asa}$, ${\rm BR}(A \to sa)$ is maximised around $\tan\beta \sim 7$.
This is because the decay rate of $A \to f \bar f$ is minimised in this region. 
For small ($\lsim 2$) and large ($\gsim 30$) $\tan\beta$, $|g_{Asa}|/(246 \, \rm GeV) \gsim 1.5$ is required 
to have ${\rm BR}(A \to sa) \gsim 0.3$.

We focus on the process in which $a$ decays to two photons through higgsino loop.\footnote{
    Similar idea has been discussed \cite{SchmidtHoberg:2012yy} in the context of the 125 GeV Higgs boson. 
} 
If $a$ is pure singlet and the gauginos are decoupled, ${\rm BR}(a \to \gamma \gamma)$ 
does not depend on the higgsino mass nor the $a \tilde h^+ \tilde h^-$ coupling, and
is entirely determined by quantum numbers of higgsinos.
The branching ratios are given as
\be
{\rm BR}(a \to W^+ W^-) \approx 0.65, ~~~~
{\rm BR}(a \to Z Z) \approx 0.23, ~~~~
{\rm BR}(a \to \gamma Z) \approx 0.05, ~~~~ 
{\rm BR}(a \to \gamma \gamma) \approx 0.07.
\label{eq:BR}
\ee

\begin{figure}
    \centering   
    \includegraphics[width=0.49\textwidth]{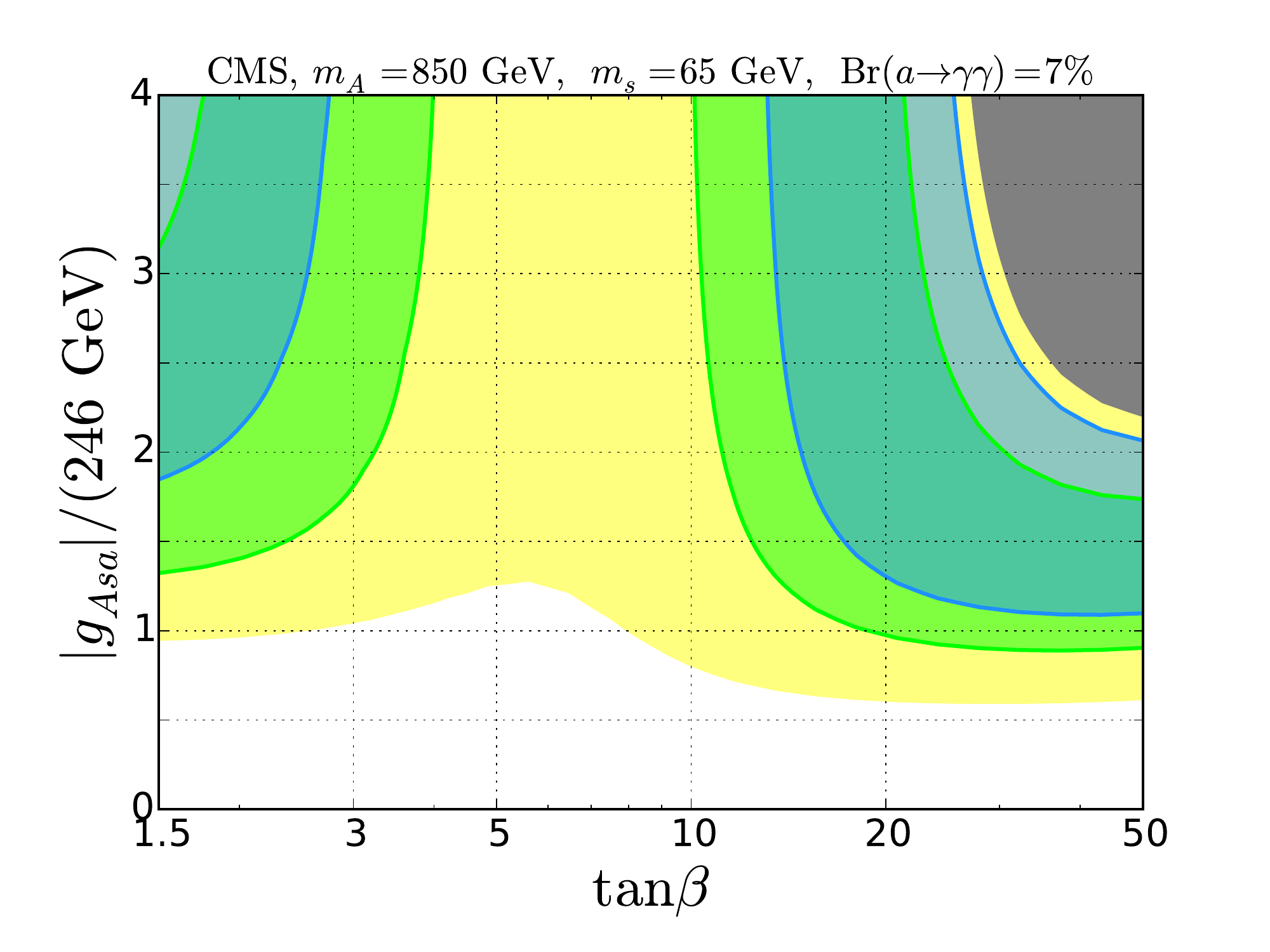}    
    \hspace{-0.5cm}
    \includegraphics[width=0.49\textwidth]{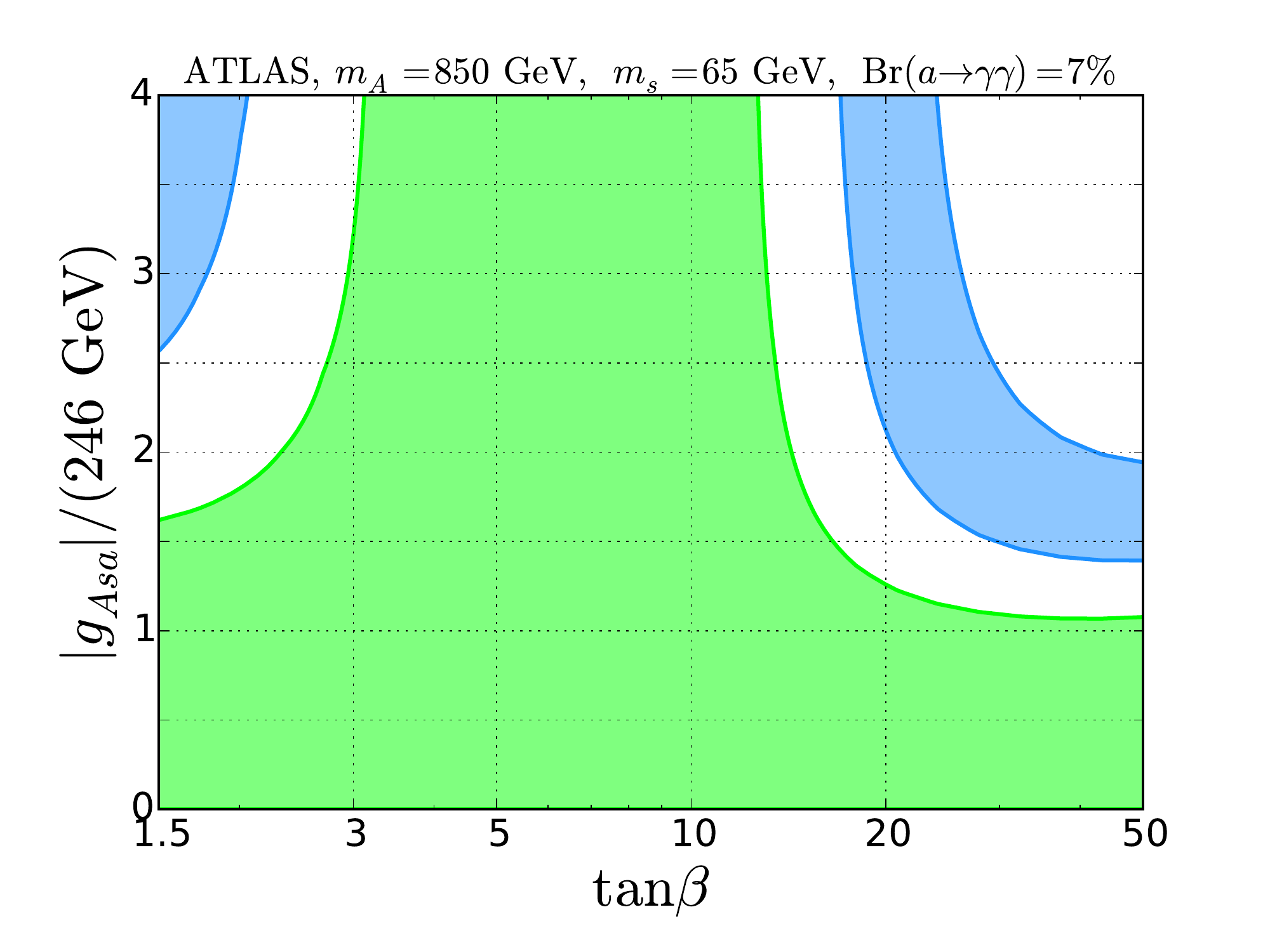}        
    \vspace{-0.4cm}
\caption{\small
{
Left: The results for the coupling $g_{Asa}$ as a function of $\tan\beta$ based on the CMS fit.
The blue (yellow) region is favoured by the 13 TeV excess at 1\,$\sigma$\,(2\,$\sigma$) level.
The green region is favoured by the  excess in the 8 TeV data at 1\,$\sigma$ level. (The blue and green regions partly overlap.) 
The grey region is beyond the  2-$\sigma$ in the 8 TeV data.  Right: The results for the coupling $g_{Asa}$ as a function of $\tan\beta$ based on 
the  fit of ref.~\cite{Franceschini:2016gxv} to the  ATLAS data. The blue region is favoured by the 13 TeV excess at 1\,$\sigma$ level.
The green region is favoured by the  excess in the 8 TeV data at 1\,$\sigma$ level. 
}
\label{fig:nomix_mA900}}
\end{figure}

We now combine the cross section and branching ratios to see if 
the model can fit the 13 TeV excess consistently with the 8 TeV data. 
Since the CMS detailed data analysis and the fit of ref.~\cite{Franceschini:2016gxv} to the ATLAS data are not on equal footing, we do not average
their results and discuss them in turn.
The results for the  coupling $g_{Asa}$ based on the CMS analysis are summarized in the  left panel of Fig.~\ref{fig:nomix_mA900}. 
The blue region is favoured by the 13 TeV excess at 1\,$\sigma$ level, $(\sigma \cdot {\rm BR})^{\rm signal}_{\rm 13 TeV} \in [2.6, \, 7.7]\,$ fb, and the yellow one by the  2\,$\sigma$ range $[0.85,\,12]$ fb.
The green region is favoured by the excess in the 8 TeV data at 1\,$\sigma$ level, 
$(\sigma \cdot {\rm BR})^{\rm signal}_{\rm 8 TeV} \in [0.31,\,1.00]$ fb.
The grey region 
corresponds to
$(\sigma \cdot {\rm BR})^{\rm signal}_{\rm 8 TeV} > 1.45 ~ {\rm fb}$
which 
is disfavoured at 2-$\sigma$ at 8 TeV.

As can be seen, there exist two favoured regions,
($a$) small ($\lsim 2$) $\tan\beta$ region and 
($b$) large ($\gsim 20$) $\tan\beta$ region. 
This is because the production cross section, $pp \to A$, is maximised for these two regions.
In the small $\tan\beta$ region $gg \to A$ via the top-quark loop dominates the production processes,
whereas $b \bar b \to A$ is dominant in the large $\tan\beta$ region.
The enhancement in the cross section compensates the slight suppression in ${\rm BR}(A \to sa)$ (see Fig.~\ref{fig:br}).
For moderate $\tan\beta$, the signal event rate cannot be large enough to be within the 1\,$\sigma$ regions due to the small cross
section
even for $|g_{Asa}|/(246\,{\rm GeV}) \gsim 1.5$ where the ${\rm BR}(A \to sa)$ is already saturated ${\rm BR}(A \to sa) \sim 1$
and increasing $g_{Asa}$ further does not help to enhance the signal rate.
As can be seen, both favoured regions require relatively large $g_{Asa}$ coupling.
For large and small $\tan\beta$ regions, the 1\,$\sigma$ region requires $|g_{Asa}|/(246\,{\rm GeV}) \gsim 1$ and 2, respectively.

In the right panel of Fig.~\ref{fig:nomix_mA900} we show the results for the  coupling  $g_{Asa}$  based on the fit of
ref.~\cite{Franceschini:2016gxv} to
the ATLAS data.  
We see that the tension between the 13 and 8 TeV data does not disappear  even with the cascade decay topology, where the primary object has the mass
of 850 GeV, and remains at the level of approximately 2\,$\sigma$.
\footnote{ In the December ATLAS note \cite{ATLAS-CONF-2015-081}, it is stated that  the  8 and 13 TeV data sets, interpreted as  a
narrow resonance with mass of 750 GeV and produced from $gg$ initial state, are compatible to each other at  $2.2\sigma$. No
update for this number has been given after Moriond conference  and one cannot infer it from the fit
of ref.~\cite{Franceschini:2016gxv}.  We note that for 850 GeV resonance produced from $b\bar{b}$ initial state the increase of the
cross-section from 8 to 13 TeV is about $40\%$ bigger than that for 750 GeV resonance produced from $gg$ initial state. In the right panel of
Fig.~\ref{fig:nomix_mA900} we see that, indeed,  the compatibility
between the fits of  ref.~\cite{Franceschini:2016gxv}  to 8 and 13 TeV data sets is  somewhat better at large
$\tan\beta$  than at its small values.}
The excess at 13 TeV requires, at 1\,$\sigma$, somewhat larger values of the coupling $g_{Asa}$.

In the simplified framework discussed so far, the dominant decay mode of $s$ becomes $s \to \gamma \gamma$
because other gauge boson final states are not kinematically allowed.
This will cause a strong tension with the fact that ATLAS and CMS did not observe 
extra photons other than the diphoton excess with $m_{\gamma \gamma} \simeq 750$ GeV.
However, this problem can be easily circumvented by introducing a mixing between $s$ and $H$.
With this mixing $s$ will dominantly decay into $b \bar b$.

\section{Realisation in NMSSM}
\label{sec:nmssm}

The superpotential and soft SUSY breaking Lagrangian of the NMSSM are given by (c.f.~\cite{Ellwanger:2009dp})
\be
W = W_{\rm MSSM} + \lambda S H_u H_d + \xi_F S + \frac{1}{2} \mu^\prime S^2 +\frac{ \kappa}{3} S^3 \,,
\ee
\be
- {\cal L}_{\rm soft} = - {\cal L}_{\rm soft}^{\rm MSSM} + m_S^2 |S|^2 +
\Big[ \lambda A_\lambda S H_u H_d + \frac{1}{3} \kappa A_\kappa S^3 + \frac{1}{2} m^{\prime 2}_3 S^2 + \xi_S S + {\rm h.c.} \Big] \,, 
\ee
where we assume all couplings are real.\footnote{
    We use general NMSSM Lagrangian without imposing $Z_3$ or scale invariance.
    This version of NMSSM has various phenomenological advantages.  See e.g.~\cite{Lee:2011dya,Ross:2011xv}. 
}
Notice that the MSSM $\mu$-term, $W_{\rm MSSM} \supset \mu_{\rm MSSM} H_u H_d$,
can be removed by redefining $S$ by a constant shift.
We fix $S$ in this way, hence $\mu_{\rm MSSM} = 0$.
We first rotate the doublet Higgs bosons $H_u$ and $H_d$ by the angle $\beta$
and define the new field basis
\ba
\hat H = \sin\beta H_{dR}^0 - \cos\beta H_{uR}^0 
\,, ~~ 
\hat h = \cos\beta H_{dR}^0 + \sin\beta H_{uR}^0 
\,, ~~
\hat{s}=S_R \,,
\\
\hat A = \sin\beta H_{dI}^0 + \cos\beta H_{uI}^0
\,, ~~~ 
\hat G = \cos\beta H_{dI}^0 - \sin\beta H_{uI}^0 
\,, ~~~
\hat{a}=S_I \,.
\ea
By this rotation, $\hat H$ does not have the vacuum expectation value, 
and $\hat G$ becomes the Goldstone boson eaten by $Z$.
The scalar mass eigenstates, denoted by $h_i$ 
(with $h_i=h,H,s$ where $h$ is the SM-like Higgs), 
are expressed in terms of the hatted fields with the help of the 
diagonalization matrix $\tilde{S}$:
\begin{equation}
h_i
=\tilde{S}_{h_i\hat{h}}\hat{h}
+\tilde{S}_{h_i\hat{H}}\hat{H}
+\tilde{S}_{h_i\hat{s}}\hat{s}
\,.
\label{tilde_S}
\end{equation}
The pseudoscalar mass eigenstates, $A$ and $a$, are related to the 
hatted fields, $\hat{A}$ and $\hat{a}$, by a rotation by angle $\theta_{Aa}$.

The $\hat A$-$\hat s$-$\hat a$ interaction is given by the F-term of $S$ as
$| \frac{\partial W}{\partial S}|^2 \supset \lambda \kappa 
H_u H_d S^* S^* \supset - v_{\rm SM} \lambda \kappa \hat A \hat s \hat a$,
where $v_{\rm SM} = 246$ GeV. In the previous section we mentioned that 
one should allow the $\hat{H}$-$\hat{s}$ mixing
in order to suppress unwanted $s \to \gamma \gamma$ decay.
Neglecting $A$-$a$ mixing, the coupling $g_{A s a}$ is given as
\be
g_{A s a} / v_{\rm SM} 
= - \lambda \kappa \tilde{S}_{s\hat{s}} \,.
\label{eq:gAsa}
\ee

\begin{figure}
    \centering 
    \includegraphics[width=0.6\textwidth]{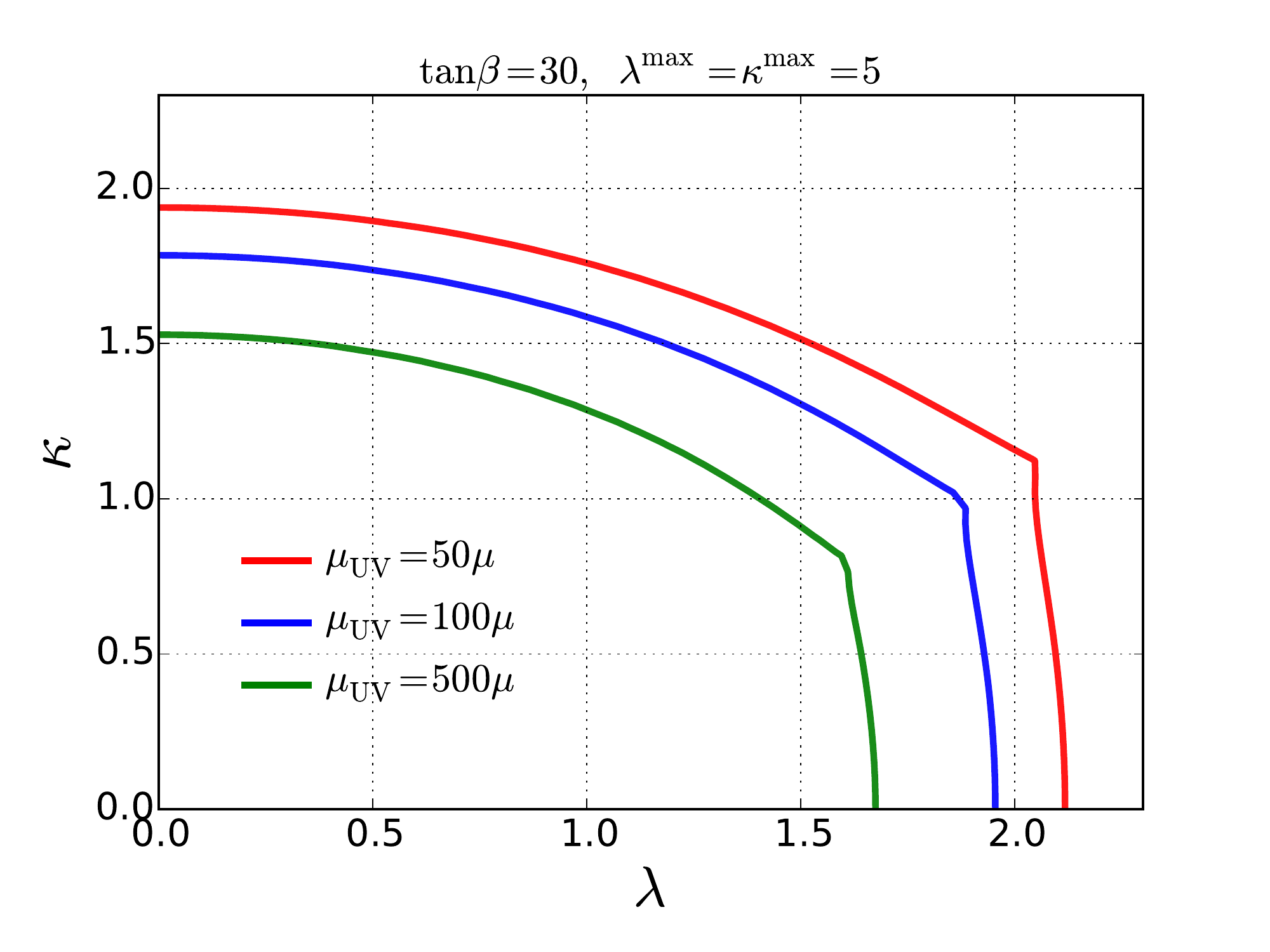}
    \vspace{-0.4cm}
\caption{
The limit for the Landau pole constraint.
At the green (blue and red) contour
${\rm max}[ \lambda(\mu_{\rm UV}), \kappa(\mu_{\rm UV}) ] = 5$,
where $\mu_{\rm UV} = 500 \mu$, ($100 \mu$, $50 \mu$).
\label{fig:landau}}
\end{figure}

In the previous section we have shown that 
$|g_{A s a} / v_{\rm SM}| \gsim 1$~(2) is required 
for $\tan\beta \gsim 20$ ($\lsim 2$)
(See Fig.~\ref{fig:nomix_mA900}.).
Clearly,  one needs the product $|\lambda \kappa| \gsim 1$~(2) 
for large (small) $\tan\beta$ to explain the excess.
Such large values of $\lambda$ and/or $\kappa$ indicate the Landau pole 
at the scale $\mu_{\rm UV}$ much below the GUT scale.
In Fig.~\ref{fig:landau} we show the constraint from the Landau pole.
If  our topology is responsible for the observed diphoton excess, 
this indicates the existence of the UV cut-off typically  of the order of  100 TeV.

Dropping the Goldstone mode, the entries of the mass matrix for the pseudo-scalar sector $(A, a)$ are given by
\ba
M^2_{\hat A \hat A} &=& \frac{2 (\mu B_{\rm eff} + \hat m_3^2) }{\sin 2\beta} + \Delta^2_{AA},
\\
M^2_{\hat a \hat a} &=& 
\frac{1}{v_s} 
\left[ \frac{\lambda v_{\rm SM}^2 \sin 2 \beta}{4}(B_{\rm eff} + \mu^\prime) - \xi_F \mu^\prime - \xi_S \right] 
\nonumber \\ &~&
+ \kappa \left[ \frac{3 \lambda v_{\rm SM}^2 \sin 2 \beta}{4} -4 \xi_F \right] 
- 2 m^{\prime 2}_S 
- \kappa v_s (3 A_\kappa + \mu^\prime) + \Delta^2_{aa},
\\
M^2_{\hat A \hat a} &=& \frac{\lambda v_{\rm SM}}{\sqrt{2}} ( A_\lambda - 2 \kappa v_s - \mu^\prime ) + \Delta^2_{Aa}
\,,
\ea
where $B_{\rm eff} \equiv A_\lambda + \kappa v_s$ and $\hat m_3^2 \equiv m_3^2 + \lambda (\mu^\prime v_s + \xi_F)$.
The $m_3^2$ is the soft breaking mass term ${\cal L}_{\rm soft}^{\rm MSSM} \supset m_3^2 H_u H_d$, 
$v_s \equiv \langle s \rangle \equiv \mu/\lambda$
and $\Delta^2_{AA/aa/Aa}$ are the radiative corrections.

The mixing between $A$ and $a$ is determined by
\be
\sin 2 \theta_{Aa} = \frac{2 M^2_{\hat A \hat a} }{m_A^2 - m_a^2} \simeq \frac{\lambda ( A_\lambda - 2 \kappa v_s - \mu^\prime )}{460 \,{\rm GeV}} 
+ \frac{\Delta^2_{Aa}}{(283 \, {\rm GeV})^2} \,,
\label{eq:mix_Aa}
\ee
where we used $m_A = 850$ GeV, $m_a = 750$ GeV.
This mixing strongly affects the ${\rm BR}(a \to \gamma \gamma )$
because it introduces $a \to b \bar b$ and $t \bar t$ modes through the mixing.
The reduction of the signal strength can be parameterized by $r$ as
\be
(\sigma \cdot {\rm BR})^{\rm signal} = r \cdot (\sigma \cdot {\rm BR})^{\rm signal}_{\rm pure}  \,.
\ee
For $|\sin \theta_{Aa}| \ll 1$, $r$ can be written as
\be
r = \frac{\cos^2\theta_{Aa} \Gamma^a_{VV}}{\sin^2\theta_{Aa} \Gamma^A_{f \bar f} + \cos^2\theta_{Aa} \Gamma^a_{VV} }
\ee
where $\Gamma^A_{f \bar f}$ is the sum of the partial decay rates in Eq.~\eqref{eq:Aff} at $m_A = 750$ GeV
and $\Gamma^a_{VV}$ is the sum of the partial decay rates of the pure state $a$ into $W^+ W^-$, $ZZ$, $Z \gamma$ and $\gamma \gamma$, which can be written as
\be
\Gamma^a_{VV} = |\lambda|^2 f(m_{\tilde h}) \,.
\ee
The factor $|\lambda|^2$ can be understood because $a \tilde h \tilde h$ coupling is given by $\frac{\lambda}{\sqrt{2}}$.
The $f(m_{\tilde h})$ is obtained from the higgsino loop diagram and we find $f(m_{\tilde h}) \simeq 1.5 \cdot 10^{-2}$ GeV for $m_{\tilde h} = |\mu| \simeq 375$ GeV. 
The condition $r \gsim 0.5$ can be translated as 
\be
|\tan \theta_{Aa}| \lsim \Big[ \frac{|\lambda|^2 f(m_{\tilde h})}{\Gamma^A_{f \bar f}} \Big]^{1/2} 
\sim 0.03 \, |\lambda|
\ee
for large ($\gsim 10$) or small ($\lsim 2$) $\tan\beta$.
This puts a strong constraint on the parameters appearing in Eq.~\eqref{eq:mix_Aa}.

In the scalar sector 
($\hat{H}$, $\hat{h}$, $\hat{s}$),
the elements of the mass matrix are given by
\ba
M^2_{\hat H \hat H} &=& M^2_{AA} + (m_Z^2 
- \frac{ \lambda^2}{2} v_{\rm SM}^2) \sin^2 2 \beta 
+ \Delta^2_{HH},
\\
M^2_{\hat h \hat h} &=& m_Z^2 \cos^2 2 \beta + \frac{ \lambda^2}{2} v_{\rm SM}^2 \sin^2 2 \beta + (\delta m_h^2)^{\rm rad} 
+ \Delta^2_{hh},
\\
M^2_{\hat s \hat s} &=& \kappa v_s (4 \kappa v_s + A_\kappa + 3 \mu^\prime) 
+ \frac{1}{v_s} 
\Big[ 
\frac{\lambda v_{\rm SM}^2 \sin 2 \beta}{4} (A_\lambda + \mu^\prime) - (\mu^\prime \xi_F + \xi_S)
\Big] 
+ \Delta^2_{s s},
\\
M^2_{\hat H \hat h} &=& \frac{1}{2}(m_Z^2 - \frac{ \lambda^2}{2} v_{\rm SM}^2) \sin 4 \beta + \Delta^2_{Hh},
\label{M2Hh}
\\
M^2_{\hat H \hat s} &=& \frac{ \lambda}{\sqrt{2}} v_{\rm SM} \Lambda \cos 2 \beta + \Delta^2_{Hs},
\label{M2Hs}
\\
M^2_{\hat h \hat s} &=& \frac{\lambda}{\sqrt{2}} v_{\rm SM} ( 2 \mu - \Lambda \sin 2 \beta ) + \Delta^2_{hs} \,,
\label{M2hs}
\ea
where $\Lambda \equiv B_{\rm eff} + \kappa v_s + \mu^\prime = A_\lambda + 2 \kappa v_s + \mu^\prime$
and $(\delta m_h^2)^{\rm rad}$ is  the radiative correction induced by the stop loop.
Typically, for large $\tan\beta$ this scenario requires heavy stops ($m_{\tilde t} \sim {\cal O}(10)$ TeV) depending on the
size of the stop mixing parameter $X_t$ in order to achieve $m_h = 125$ GeV.
The $\Delta^2_{HH/hh/ss/Hh/Hs/hs}$ are  the radiative corrections contributing  to  the NMSSM Higgs boson  mass matrices.

The elements of the diagonalization matrix $\tilde{S}$ 
must respect various phenomenological constraints.
The LEP limit on the $e^+ e^- \to Z^* \to Z s$ $(s \to b \bar b)$ process
for the 65 GeV scalar gives the bound 
$\tilde{S}_{s\hat{h}} \cdot {\rm BR}(s \to b \bar b) \lsim 0.16$ 
\cite{LEP:2006},
where ${\rm BR}(s \to b \bar b)$ depends in principle on 
$\tilde{S}_{s\hat{H}}$ mixing and $\tan\beta$ 
\cite{Badziak:2013bda}.
The measurements of the properties of the SM-like Higgs boson 
at the LHC also give constraints on the mixing angles.
The deviation of its coupling to the gauge bosons is now constrained 
up to $\sim 20$\,\% at 95\,\% CL
\cite{Aad:2015gba,Khachatryan:2014jba}.
This translates into the constraint on the 
$\tilde{S}$ entries 
as $\tilde{S}_{h\hat{H}}, \tilde{S}_{h\hat{s}} \lsim 0.2$.

In the parameter space relevant for our model, the 
elements $\tilde{S}_{s\hat{H}}$ and $\tilde{S}_{H\hat{s}}$  
remain unconstrained  and may be large.
Neglecting  the  small mixing elements 
they may be approximated by
\begin{equation}
\tilde{S}_{s\hat{H}} \simeq \sin\theta_{sH} \simeq -\tilde{S}_{H\hat{s}}
\,,
\end{equation}
where for future convenience we introduced the mixing angle $\theta_{sH}$ 
satisfying
\begin{equation}
\sin2\theta_{sH}=
\frac{2 M^2_{Hs}}{m_s^2 - m_H^2} 
\simeq -\frac{\lambda \Lambda \cos 2 \beta}
{2 \, {\rm TeV}} - \frac{\Delta^2_{Hs}}{(600 \, {\rm GeV})^2}\,,
\end{equation}
In the last equality we have used $m_H = 850$ GeV, $m_s = 65$ GeV.
The two small off-diagonal entries of $\tilde{S}$ may be approximated as 
follows
\begin{equation}
\tilde{S}_{s\hat{h}} 
\simeq
\frac{\cos\theta_{sH} M^2_{\hat{h}\hat{s}} + \sin\theta_{sH} M^2_{\hat{H}\hat{h}}}
{m_s^2 - m_h^2}\,,
\qquad 
\tilde{S}_{H\hat{h}}  
\simeq 
\frac{\cos\theta_{sH} M^2_{\hat{H}\hat{h}} - \sin\theta_{sH} M^2_{\hat{h}\hat{s}}}
{m_H^2 - m_h^2}
\,.
\label{SshSHh}
\end{equation}
The elements $\tilde{S}_{h\hat{H}}$ and $\tilde{S}_{h\hat{s}}$ are related to the 
above ones by orthogonality of $\tilde{S}$:
\begin{equation}
\tilde{S}_{h\hat{s}} 
\simeq
-\cos\theta_{sH} \tilde{S}_{s\hat{h}} + \sin\theta_{sH} \tilde{S}_{H\hat{h}} 
\,,
\qquad
\tilde{S}_{h\hat{H}} 
\simeq 
-\cos\theta_{sH} \tilde{S}_{H\hat{h}} - \sin\theta_{sH} \tilde{S}_{s\hat{h}} 
\,.
\label{ShsShH}
\end{equation}

Clearly, the values of the Higgs boson masses and the constraints on the mixing angles
would select some regions of the NMSSM parameter space. 
However, the complexity of the NMSSM Higgs potential 
make a full quantitative analysis of our scenario, with radiative corrections included, challenging and premature.
Merely for the illustration purpose, we attempt to find the NMSSM parameters 
that satisfy the above conditions using approximate forms of the 1-loop radiative corrections.
Some attention has to be paid to the magnitude of the radiative corrections.
Indeed, we note that some of the 1-loop radiative correction terms are proportional to 
the 3rd power of $\lambda$ or $\kappa$ and can be as large as the tree level terms 
for $|\lambda|$, $|\kappa| \gsim 1$ \cite{Ellwanger:2005fh}.
The 2-loop corrections may also be large \cite{Goodsell:2014pla} in this region.\footnote{
  For instance, a brute force parameter scan using numerical tools that 
include such corrections
  is computationally very expensive since one has to find a  narrow region 
  where the mixing parameters, $\sin\theta_{Aa}$ and $\tilde{S}_{s\hat{h}}$, 
are  small. 
}
For large $\lambda$ and $\kappa$, 
neglecting the corrections proportional to the gauge and Yukawa couplings,
the leading terms of the radiative corrections 
to the off-diagonal mass matrix elements are given by\footnote
{We applied the loop corrections from ref.~\cite{Ellwanger:2005fh} modified 
by the $Z_3$ non-invariant contributions.
}
\ba
\Delta^2_{Hs}&=&
\frac{\kappa v_{\rm SM}\mu}{8\sqrt{2}\pi^2}\,
\left(
2\lambda^2 L_\mu +2\kappa^2 L_\nu -3\lambda^2 L_{\mu\nu}
\right)
\cos(2\beta)\,,
\\
\Delta^2_{hs}&=&
\frac{\lambda v_{\rm SM}\mu}{8\sqrt{2}\pi^2}\,
\left(
2\lambda^2 L_\mu + 2\kappa^2 L_\nu 
-\left(\lambda^2+8\kappa^2\right) L_{\mu\nu}
\right)
-\Delta^2_{Hs}\tan(2\beta)
\,,\\
\Delta^2_{Aa}&=&
\Delta^2_{Hh}\,\,=\,\,0\,,
\ea
where 
\begin{equation}
L_\mu=\ln\left(\frac{\mu^2}{M_Z^2}\right)
,\qquad
L_\nu=\ln\left(\frac{(2\kappa v_s+\mu')^2}{M_Z^2}\right)
,\qquad
L_{\mu\nu}=\ln\left(\frac{\max(\mu^2,(2\kappa v_s+\mu')^2)}{M_Z^2}\right)
.
\end{equation}
It is easy to find solutions for the parameters of the model satisfying 
the constraints  $m_H = m_A = 850$ GeV, $m_s=65$ GeV , $\mu=375$~GeV,
vanishing $Aa$ mixing ($\theta_{Aa}=0$) and small $\tilde{S}_{s\hat{h}}$.
We used the following procedure:
The scalar mass squared matrix has 6 independent parameters. 
We choose them as 3 eigenvalues ($m_h^2$, $m_H^2$, $m_s^2$) and 3 
off-diagonal entries of the diagonalization matrix 
($\tilde{S}_{s\hat{H}}$, $\tilde{S}_{s\hat{h}}$, $\tilde{S}_{h\hat{H}}$).
Using this parameterization we calculate the off-diagonal elements 
of the scalar mass squared matrix and compare them 
with the same elements expressed by the parameters 
of the model in eqs.~(\ref{M2Hh})-(\ref{M2hs}). One of the parameters, 
$\mu'$, is fixed by the requirement of vanishing $A$-$a$ mixing:
$\mu'=A_\lambda-2\mu\kappa/\lambda$. Then, for some fixed values of the elements ($\tilde{S}_{s\hat{H}}$, $\tilde{S}_{s\hat{h}}$, $\tilde{S}_{h\hat{H}}$), we are left with the set of three 
equations for three parameters: $\lambda$, $\kappa$ and $A_\lambda$. 
In general there is a discrete set of solutions.

\begin{figure}
    \centering 
    \includegraphics[width=0.49\textwidth]{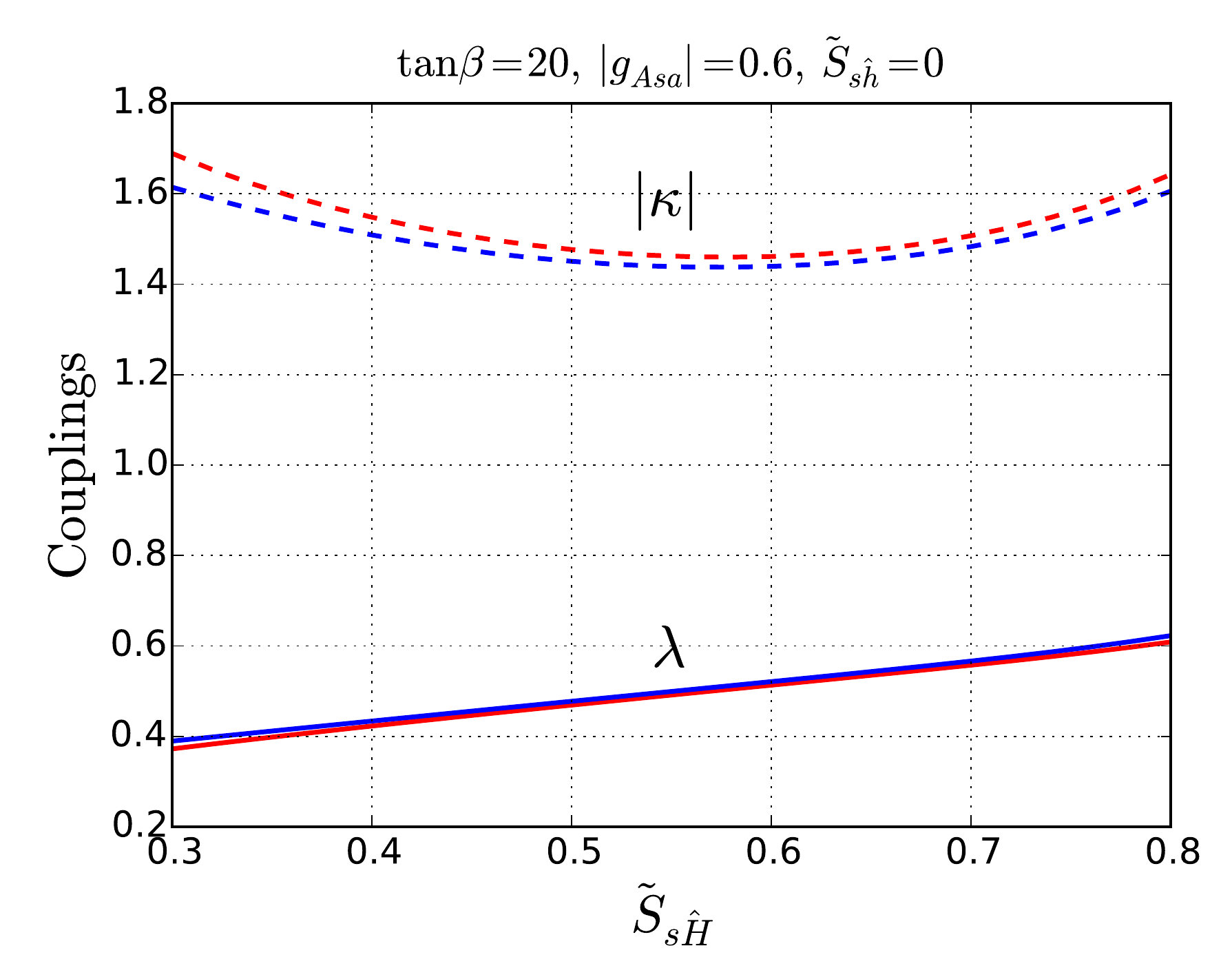}    
    \includegraphics[width=0.49\textwidth]{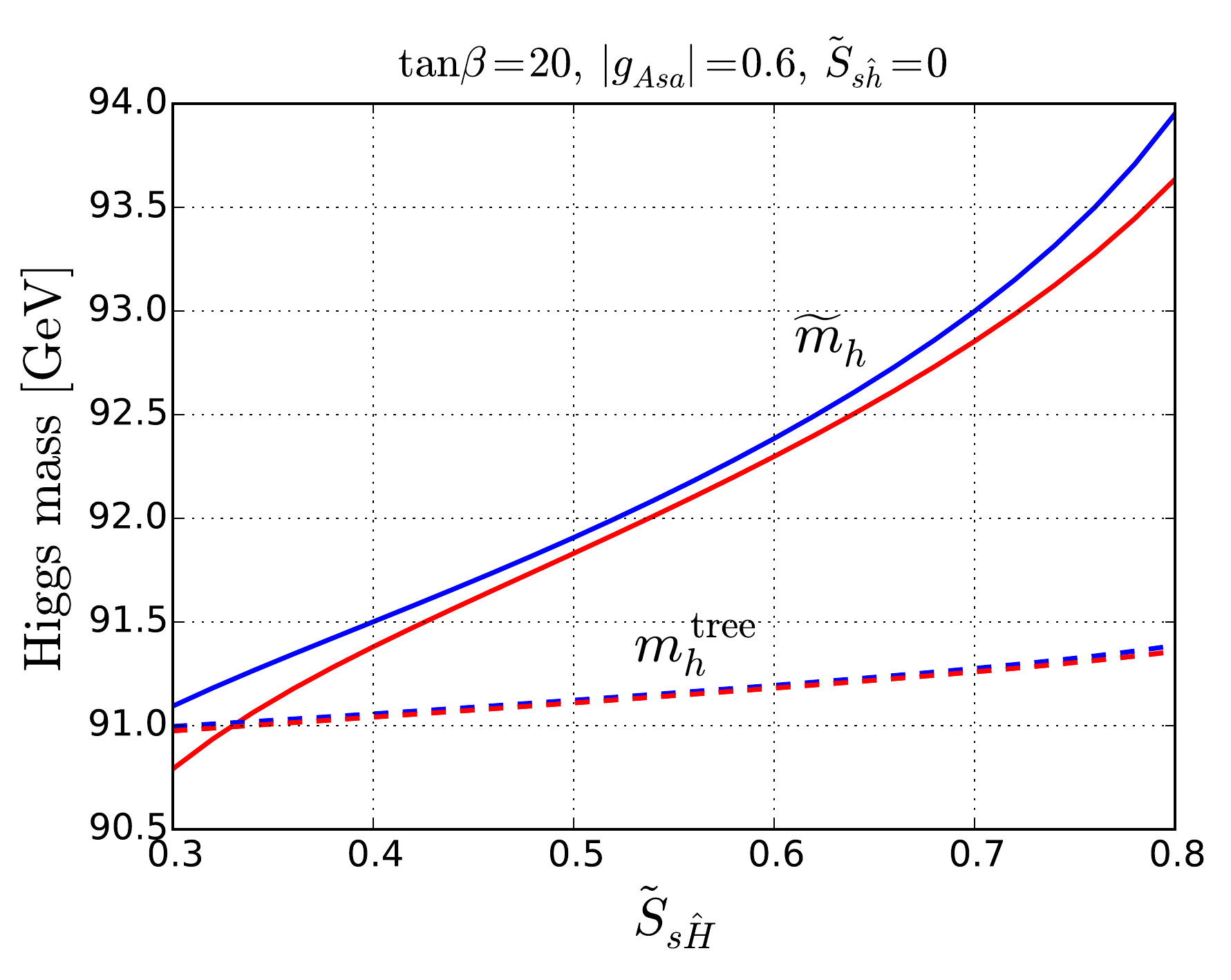}           
    \vspace{-0.4cm}
\caption{\small
Left panel: $\lambda$ (solid) and $|\kappa|$ (dashed) 
as functions of $\tilde{S}_{s\hat{H}}$.
Right panel: the SM-like Higgs boson mass at the tree level, 
$m_h^{\rm tree}$, (dashed) and with the leading 
(for large $\lambda$ and $\kappa$) loop corrections 
(but before including the radiative correction from the scalar top loop),
$\widetilde{m}_h$, (solid) as functions of $\tilde{S}_{s\hat{H}}$. 
Solutions with positive (red) and negative (blue) values of $\kappa$ are shown. 
Other parameters are fixed at: 
$\tan\beta=20$, $|g_{Asa}|=0.6$, $\tilde{S}_{s\hat{h}}=0$.
\label{fig:example}}
\end{figure}

In the actual numerical calculations we had to modify this simple 
prescription. In order to compare our results with the experimental
constraints illustrated in Fig.~\ref{fig:nomix_mA900} we were fixing 
the value of $g_{Asa}$ given by eq.~(\ref{eq:gAsa}). This fixes one 
combination of the parameters $\lambda$, $\kappa$ and $A_\lambda$. 
Thus, only two mixing elements 
(chosen to be $\tilde{S}_{s\hat{H}}$, $\tilde{S}_{s\hat{h}}$) 
remain as input for our calculations while the third one ($\tilde{S}_{h\hat{H}}$) 
is obtained as output. Numerical iteration procedures are used to 
find solutions.

One of the input mixing elements, $\tilde{S}_{s\hat{h}}$, is quite 
strongly constrained by the LEP data. Thus, after fixing the values 
of the scalar masses and $\tan\beta$, $\tilde{S}_{s\hat{H}}$ remains 
the only input quantity which may be changed in a relatively wide range.
The dependence of the results on $\tilde{S}_{s\hat{H}}$ 
is shown in Fig.~\ref{fig:example} for the example with 
$\tan\beta=20$, $\tilde{S}_{s\hat{h}}=0$ and $|g_{Asa}|=0.6$.
One can see that $\lambda$ increases with $\tilde{S}_{s\hat{H}}$ 
while $|\kappa|$ has a minimum. 
The behavior of $\lambda$ follows from the fact that for bigger 
mixing $\tilde{S}_{s\hat{H}}$ one needs bigger $M^2_{\hat{H}\hat{s}}$ which
grows with $\lambda$ (at least the tree contribution, see eq.~(\ref{M2Hs})).
Then the behavior of $\kappa$ follows from relation (\ref{eq:gAsa}). 
The leading (in $\lambda$ and $\kappa$) 
loop correction to the Higgs mass is a quite complicated function 
of the parameters. From the right panel in  Fig.~\ref{fig:example}
one can see that it may even vanish for 
some combination of $\lambda$ and $\kappa$ but generally is 
an increasing function of the input mixing parameter $\tilde{S}_{s\hat{H}}$.
Examples presented in Fig.~\ref{fig:example} (and in Table~\ref{tab1}) 
were obtained for $\tilde{S}_{s\hat{h}}=0$. We checked that the results 
do not change substantially for the  values of  
$\tilde{S}_{s\hat{h}}$ allowed by the LEP data.

A few generic examples are presented in Table~\ref{tab1}.
For large $\tan\beta$ the values of $\tilde{S}_{s\hat{H}}$ are chosen 
to give $|\kappa|$ close to the smallest possible (for a given set 
of other parameters) value in order to get the Landau pole scale 
as big as possible. For small $\tan\beta$ we have to choose much smaller 
$\tilde{S}_{s\hat{H}}$ in order to avoid huge tree level contribution 
to the Higgs mass (value of $\lambda$ increases with $\tilde{S}_{s\hat{H}}$). 
The first example in Table~\ref{tab1} shows that $\tilde{S}_{s\hat{H}}\gsim0.1$ 
can easily lead to too large $m_h^{\rm tree}$ for $\tan\beta=2$. 
The last two columns of Table~\ref{tab1} show the SM-like Higgs boson
at the tree level, $m_h^{\rm tree}$, and with the leading 
(for large $\lambda$ and $\kappa$) loop corrections, $\widetilde m_h$ 
(but before including the radiative correction from the scalar top loop). 
An interesting observation is that in the parameter range selected 
by the constraints of very small
$\hat{h}$-$\hat{s}$ and $\hat{A}$-$\hat{a}$ mixings  
the radiative corrections to the Higgs potential from the NMSSM Higgs bosons 
are  actually small, in spite of the sizable values of $\lambda$ and, 
particularly, $\kappa$. 
This is related to the fact that some of potentially large contributions 
are proportional to appropriate mixing elements and are small in the 
limit of small mixings.  
Thus, the values of $\widetilde{m}_h$ given 
in Table~\ref{tab1} are almost entirely controlled by the tree-level effects.
The mixing elements, other than
$\tilde{S}_{s\hat{H}}\approx-\tilde{S}_{H\hat{s}}$,
are small once $\tilde{S}_{s\hat{h}}$ is taken to be small 
(to fulfill the LEP constrains). 
$\tilde{S}_{H\hat{h}}$ is suppressed by 
$m_H^2$ (see eqs.~(\ref{SshSHh})) and typically is below 0.01. 
The two remaining off-diagonal elements are also small due to 
relations~(\ref{ShsShH}). 
$\tilde{S}_{h\hat{s}}\approx-\tilde{S}_{s\hat{h}}$ up to 
small corrections while $|\tilde{S}_{h\hat{H}}|<0.1$  
($<0.01$ in most cases). All these mixing elements are well below 
present experimental bounds.
The numbers given in the table 
illustrate the expected order of magnitude for the soft mass parameters 
necessary to explain the di-photon excess in our scenario and indicate 
that it will be fine-tuned at the level of 1 per mille.

\begin{table}
\begin{center}
\begin{tabular}{|c|c|c|c|c|c|c|c|c|}
\hline
$\tan\beta$ & $|g_{Asa}|$ & $\tilde{S}_{s\hat{H}}$ & $\lambda$ & $\kappa$ 
& $A_\lambda$ [TeV] & $\mu'$ [TeV] & $m_h^{\rm tree}$ [GeV] 
& $\widetilde{m}_h$ [GeV] \\
\hline
2 & 2.1 & 0.15 & 1.38 & $-$1.54 & 0.39 & 1.23 & 199 & 215 \\
\hline
2 & 1.4 & 0.05 & 0.69 & $-$2.04 & 0.41 & 2.63 & 110 & 112 \\
\hline
2 & 1.0 & 0.09 & 0.79 & $-$1.27 & 0.42 & 1.62 & 123 & 125 \\
\hline
2 & 1.0 & 0.06 & 0.62 & 1.61 & 0.27 & 1.68 & 102 & 113 \\
\hline
7 & 1.4 & 0.4 & 0.97 & $-$1.57 & 0.87 & 2.07 & 100 & 112 \\
\hline
20 & 1.3 & 0.5 & 0.80 & $-$1.88 & 1.29 & 3.05 & 92 & 96 \\
\hline
20 & 1.0 & 0.6 & 0.70 & 1.78 & 1.25 & $-$0.65 & 92 & 95 \\
\hline
20 & 0.6 & 0.6 & 0.51 & 1.46 & 1.79 & $-$0.35 & 91 & 92 \\
\hline
\end{tabular}
\end{center}
\caption{Examples of solutions with vanishing $\tilde{S}_{s\hat{h}}$
and $\theta_{Aa}$ for $m_H = m_A = 850$ GeV, $m_s=65$ GeV and $\mu=375$~GeV.  
The SM-like Higgs boson mass at the tree level, 
$m_h^{\rm tree}$, and with the leading 
(for large $\lambda$ and $\kappa$) loop corrections 
(but before including the radiative correction from the scalar top loop),
$\widetilde{m}_h$, are given in the last two columns.
Mixing elements $\tilde{S}_{h\hat{s}}$, $\tilde{S}_{h\hat{H}}$
and $\tilde{S}_{H\hat{h}}$ are at most of order 0.01 for all these 
examples.
} 
\label{tab1}
\end{table}

Finally we comment on the constraint from electroweak precision tests.
It has been pointed out \cite{Barbieri:2006bg, Franceschini:2010qz, Gherghetta:2012gb} that large values of $\lambda$ and $\tan\beta$ may introduce
a dangerous contribution from light higgsinos to the $T$-parameter \cite{Peskin:1991sw} as a consequence of violation of SU(2) custodial symmetry. 
However, generically, in the selected region,  $\lambda <1$. Moreover, ref.~\cite{Barbieri:2006bg} shows  that even for $\lambda = 2$  there are
strips  around the singlino mass parameter 
$|\mu_s| = |\mu^\prime + (\kappa/\lambda) \mu| \simeq 750 \div 800$ GeV   
where the higgsino contribution to the $T$-parameter vanishes or is negative  independently of $\tan\beta$  and weakly dependent on the value of
$\mu$. It is not difficult to find solutions with the singlino mass in the above range, as for instance the last example in Table~\ref{tab1}.
We, therefore, expect the higgsino contribution to the $T$ parameter not to be a problem for our scenario.
One can also expect some cancellation between the higgsino contribution and the contributions from NMSSM Higgs bosons.
We leave a detailed numerical analysis for future work.

\section{Conclusions}
\label{sec:concl}

We demonstrate that the plain NMSSM can explain the observed diphoton excess at $m_{\gamma \gamma} \simeq 750$ GeV as a decay of a single particle into two photons  at the price of a relatively low UV cut-off (around  100 TeV) and of a certain fine tuning of the parameters. The mechanism behind this scenario is production of a doublet-like pseudo scalar $A$, decaying into  a singlet-like pseudo scalar $a$, which subsequently decays via the vector-like higgsino loop into two photons. The predicted  width  of $a$ is very small, much below the experimental resolution.   The two-photon signal should be associated with $b$-quark jets coming from the decay $A\rightarrow as$, with $s$ decaying dominantly into a pair of $b$ quarks.  The pseudo scalar $a$  decays also into other channels with the branching ratios given by Eq.~\eqref{eq:BR}.

The topology proposed in  this paper is the only one that can explain the 750 GeV excess  in the plain
NMSSM due to a single particle decay. Another possibility for the NMSSM, recently proposed,
is to explain the observed signal by the decays of two light pseudo scalars, to two collimated
photons each. The latter interpretation could explain a broad peak at 750 GeV, if confirmed experimentally.  The width of the signal will give a crucial discrimination between different  proposed
interpretations, in particular between perturbative and non-perturbative scenarios.

\vspace{5mm}

{\bf Acknowledgments}
We thank Michael Schmidt, Kai Schmidt-Hoberg and Florian Staub for valuable comments.
MO and SP have been supported by the National Science Centre, Poland, under research grants
DEC-2014/15/B/ST2/02157,
DEC-2012/04/A/ST2/00099  and DEC-2015/18/M/ST2/00054.
MB was supported in part by the Director, Office of Science,
Office of High Energy and Nuclear Physics, of the US Department of Energy under Contract
DE-AC02-05CH11231 and by the National Science Foundation under grant
PHY-1316783. MB acknowledges support from the
Polish Ministry of Science and Higher Education (decision no.\ 1266/MOB/IV/2015/0).
SP and KS thank CERN Theory Division for its hospitality during the final work on this project.


\bibliographystyle{JHEP}
\bibliography{nmssm_diphoton}

\end{document}